\documentclass[8.5pt,twoside,twocolumn]{article}
\usepackage[super,sort&compress,comma]{natbib} 
\usepackage{amsmath}
\usepackage{amssymb} 
\usepackage{amsthm}
\usepackage{textcomp}
\usepackage{gensymb}
\usepackage{appendix}
\usepackage{booktabs}

\usepackage{float}

\usepackage{sectsty}
\usepackage{balance} 
\usepackage{color}
\usepackage{bm}
\usepackage{graphicx} 
\usepackage{lastpage}
\usepackage[format=plain,justification=raggedright,singlelinecheck=false,font=small,labelfont=bf,labelsep=space]{caption} 
\usepackage{fancyhdr}
\definecolor{dgreen}{RGB}{0,127,0}
\definecolor{calmblue}{RGB}{70,130,180}

\begin{document}

\thispagestyle{plain}
\fancypagestyle{plain}{
\fancyhead[L]{\includegraphics[height=8pt]{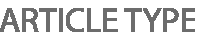}}
\fancyhead[C]{\hspace{-1cm}\includegraphics[height=20pt]{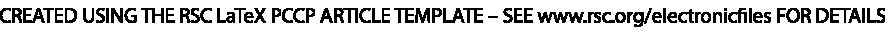}}
\fancyhead[R]{\includegraphics[height=10pt]{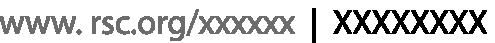}\vspace{-0.2cm}}
\renewcommand{\headrulewidth}{1pt}}
\renewcommand{\thefootnote}{\fnsymbol{footnote}}
\renewcommand\footnoterule{\vspace*{1pt}%
\hrule width 3.4in height 0.4pt \vspace*{5pt}} 
\setcounter{secnumdepth}{5}

\makeatletter 
\def\subsubsection{\@startsection{subsubsection}{3}{10pt}{-1.25ex plus -1ex minus -.1ex}{0ex plus 0ex}{\normalsize\bf}} 
\def\paragraph{\@startsection{paragraph}{4}{10pt}{-1.25ex plus -1ex minus -.1ex}{0ex plus 0ex}{\normalsize\textit}} 
\renewcommand\@biblabel[1]{#1}            
\renewcommand\@makefntext[1]%
{\noindent\makebox[0pt][r]{\@thefnmark\,}#1}
\makeatother 
\renewcommand{\figurename}{\small{Fig.}~}
\sectionfont{\large}
\subsectionfont{\normalsize} 

\fancyfoot{}
\fancyfoot[LO,RE]{\vspace{-7pt}\includegraphics[height=9pt]{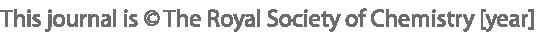}}
\fancyfoot[CO]{\vspace{-7.2pt}\hspace{12.2cm}\includegraphics{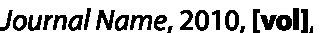}}
\fancyfoot[CE]{\vspace{-7.5pt}\hspace{-13.5cm}\includegraphics{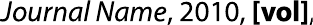}}
\fancyfoot[RO]{\footnotesize{\sffamily{1--\pageref{LastPage} ~\textbar  \hspace{2pt}\thepage}}}
\fancyfoot[LE]{\footnotesize{\sffamily{\thepage~\textbar\hspace{3.45cm} 1--\pageref{LastPage}}}}
\fancyhead{}
\renewcommand{\headrulewidth}{1pt} 
\renewcommand{\footrulewidth}{1pt}
\setlength{\arrayrulewidth}{1pt}
\setlength{\columnsep}{6.5mm}
\setlength\bibsep{1pt}

\twocolumn[
  \begin{@twocolumnfalse}
  	\noindent\LARGE{\textbf{A Single Nucleotide Resolution Model for Large-Scale Simulations of Double Stranded DNA}}

\vspace{0.6cm}

\noindent\large{\textbf{Y. A. G. Fosado,\textit{$^{a}$} D. Michieletto,\textit{$^{a}$} Jim Allan,\textit{$^{b}$} Chris Brackley,\textit{$^{a}$} O. Henrich\textit{$^{a,c}$}, D. Marenduzzo$^{\ast}$\textit{$^{a}$}}}\vspace{0.5cm}

\noindent\textit{\small{\textbf{Received Xth XXXXXXXXXX 20XX, Accepted Xth XXXXXXXXX 20XX\newline
First published on the web Xth XXXXXXXXXX 200X}}}

\noindent \textbf{\small{DOI: 10.1039/b000000x}}
\vspace{0.6cm}

\noindent \normalsize{\textbf{The computational modelling of DNA is becoming crucial in light of new advances in DNA nano-technology, single-molecule experiments and \emph{in vivo} DNA tampering. Here we present a mesoscopic model for double stranded DNA (dsDNA) at the single nucleotide level which retains the characteristic helical structure, while being able to simulate large molecules -- up to a million base pairs -- for time-scales which are relevant to physiological processes. This is made possible by an efficient and highly-parallelised implementation of the model which we discuss here. We compare the behaviour of our model with single molecule experiments where dsDNA is manipulated by external forces or torques. We also present some results on the kinetics of denaturation of linear DNA. }}
\vspace{0.5cm}
 \end{@twocolumnfalse}]

\section{Introduction}
\footnotetext{\textit{$^{a}$ School of Physics and Astronomy, University of Edinburgh, Peter Guthrie Tait Road, Edinburgh EH9 3FD, Scotland, United Kingdom.}}
\footnotetext{\textit{$^{b}$ Institute of Genetics and Molecular Medicine, MRC Human Genetics Unit, University of Edinburgh, Western General Hospital, Crewe Road, Edinburgh EH4 2XU}}
\footnotetext{\textit{$^{c}$ EPCC, University of Edinburgh, Peter Guthrie Tait Road, Edinburgh EH9 3FD, Scotland, United Kingdom.}}
\footnotetext{\textit{$^{\ast}$ Corresponding author.}}

Since the discovery of the structure of the deoxyribonucleic acid (DNA)~\cite{Crick1953,Wilkins1953,Franklin1953a}, the geometry of the double-helix and its topological implications have engaged and fascinated the scientific community~\cite{Calladine1997,Bates2005}. It is becoming more and more evident that not only is the genetic information encoded in the DNA sequence of primary importance, but also that changes in its three-dimensional structure can alter crucial biological functions, such as gene expression and replication~\cite{Cavalli2013,Brackley2013a,Cook2009,Cook2002,Alberts2014}. At the same time, the rapid improvement of techniques using DNA functionalised colloids~\cite{Leunissen2009,DiMichele2013}, DNA-origami~\cite{Rothemund2006} and, more generally, supra-molecular DNA assembly~\cite{McLaughlin2011} is setting new standards for DNA-based nano-technology. This has far-reaching applications, ranging from materials science (to create new DNA-based and possibly biomimetic materials), to medicine (to be used in, e.g., gene-therapy and drug delivery).

In light of this, the formulation of accurate theoretical and computational models that can efficiently capture the behaviour of DNA, either \emph{in vivo} or \emph{in vitro}, is of great importance in order to understand a number of outstanding biological problems, and also to assist the advance of DNA-based nanotechnology.
 
Several fully atomistic models for double-stranded (ds) DNA are available in the literature~\cite{CheathamIII2004360, doi:10.1021/ct900200k, Orozco2008185}. While these give an accurate description of the dynamics of DNA molecules and their interaction with single proteins, the complexity of the all-atom approach places severe limits on the size (up to about a hundred base-pairs) and time scales (of the order of $\mu$s) which can be probed~\cite{DAnnessa2014}. Coarse-graining, where large collections of atoms or molecules are represented by single units, allows larger systems to be simulated for longer at the expense of molecular detail. One of the most challenging aspects in designing a computational model is to retain the key microscopic details necessary to answer a given question while ``trimming'' the rest. 
At the large scale limit, entire eukaryotic chromosomes can be modelled using simple bead-and-spring polymer models~\cite{Brackley2013a,Michieletto2016drosophila}, where each monomer can represent up to $3000$ base-pairs (bp) and the simulated time can reach time-scales spanning minutes~\cite{Michieletto2016drosophila} or even days~\cite{Rosa2008}; similar chains of beads can also be used to model naked DNA, though clearly such an approach neglects microscopic details such as the base-pair specificity or the double-stranded structure. While in some cases these models can still capture the essential physics\cite{Michieletto2015}, in others they are only a crude approximation of the real systems.  
Several successful mesoscopic models have recently been proposed which aim at bridging the gap between the ``all-atom'' and ``bead-spring'' limits~\cite{Ouldridge2011a,Hinckley2013,DePablo2011,Nomidis2016}. Nevertheless, a coarse grained model able to retain the necessary physical microscopic details while allowing simulations of the several tens or hundreds of kilo-base pairs that would be needed to address many biologically relevant questions, is still currently needed.

Examples of biological processes for which such a mesoscopic approach would be highly valuable can be classified in two broad categories: processes where DNA is mechanically manipulated by enzymatic machines (for example during replication or transcription which require opening of the double-helical structure), or processes where interactions between DNA and proteins depend more subtly on the topological and geometrical properties of the double-helix. An example of the latter class of problems is the so-called ``linking number paradox'', where it has been observed that the unbinding of DNA from a nucleosome releases only one unit of writhe, rather than the 1.7 which were stored~\cite{Prunell1998,Hayes1990}; the resolution of the paradox is that the nucleosome also stores some twist (the terms twist and writhe are explained below). To complicate the picture even more, there are several proteins which operate to alter the DNA topology, whose collective actions may sometimes trigger complex feedback mechanisms that are crucial for biological functions~\cite{Ding2014,Brackley_supercoil}. For a model to be applicable to such problems, it must possess both a good accuracy in mimicking the geometry of the double-helix, and the ability to consider long molecules on which many proteins may act simultaneously, so that cooperative effects can be investigated.

Motivated by this goal, in this paper we introduce a single nucleotide resolution coarse-grained model for dsDNA which retains several biologically-relevant DNA features, while being capable of delivering large-scale simulations. The model is implemented in the LAMMPS molecular dynamics engine~\cite{LAMMPS} which allows us to comfortably study molecules on the order of thousands of bp (kbp). Because the code is fully parallel and highly scalable, it is portable to supercomputers to reach the length and time scales needed for some of the biological applications just mentioned. 
The scope of this work is to present the construction of the model, starting from the known geometry of DNA~\cite{Calladine1997} (Sec.~\ref{sec:Model}), and to discuss the validation of its main physical features, \emph{i.e.} helical pitch, persistence length and torsional rigidity  (Sec.~\ref{sec:Validation}). These properties are traditionally addressed via single-molecule experiments~\cite{Ritort2006a,Bryant2003} \emph{in vitro}, and we here provide an indirect validation via simulated single-molecule experiments, obtaining a remarkably good agreement with the experimentally observed values (Sec.~\ref{sec:SingleMolExp}). Finally, we present an application of this model to the dynamics of DNA denaturation, and discuss further future applications. These range from the study of DNA denaturation to that of supercoil dynamics in the presence of topological proteins (Sec.~\ref{sec:Discussion}). The flexibility of the model and the scalability provided by the LAMMPS engine means it provides a solid framework on which to base further studies of the topological properties of DNA and DNA-protein interactions.

\begin{figure}[!th]
	\begin{center}
		\includegraphics[width=0.5\textwidth]{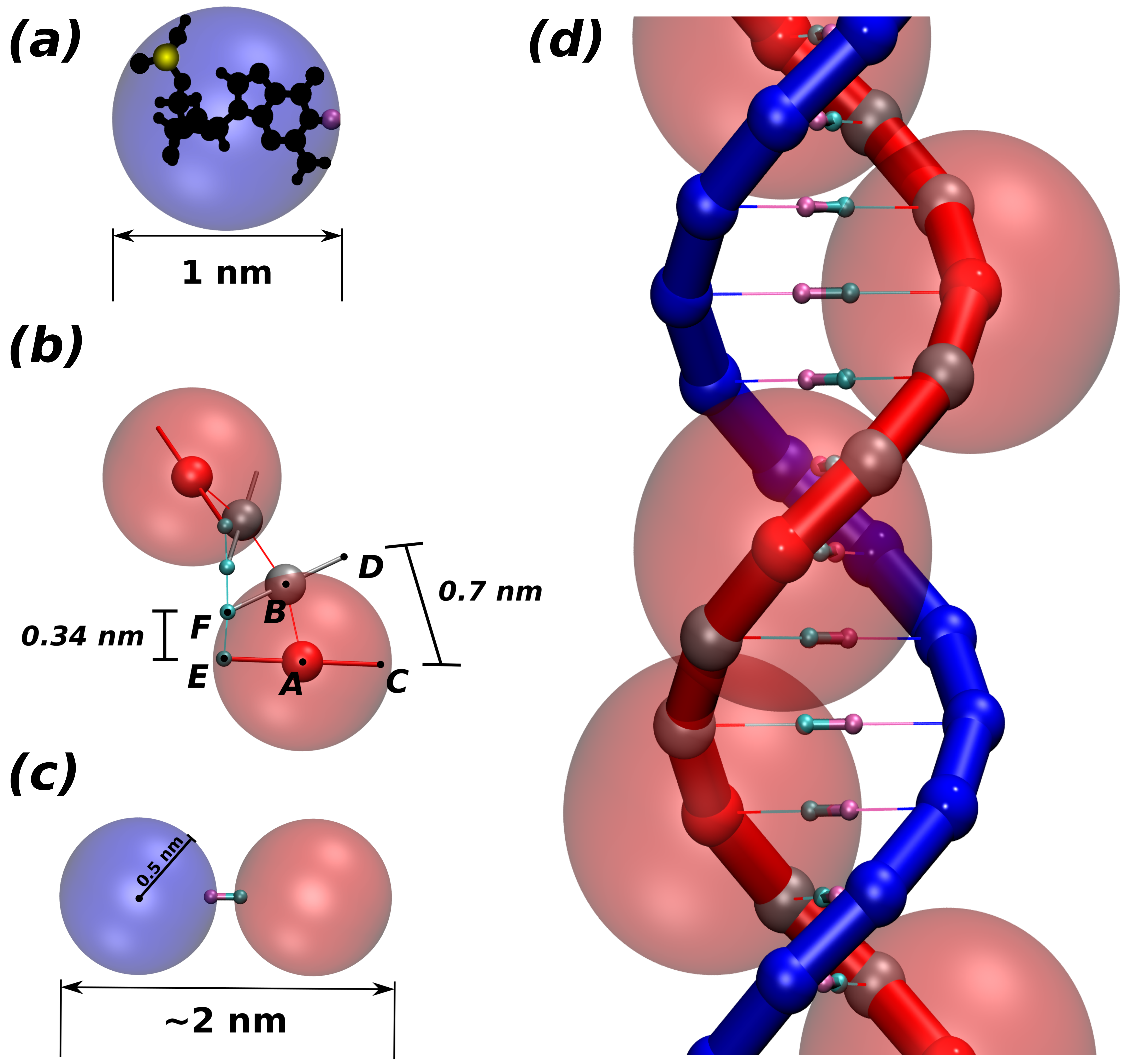}
	\end{center}
	\caption{\textbf{(a)} The level of coarse-graining of the model is here summarised by encapsulating the atoms forming one nucleotide into one bead-patch complex. The small yellow sphere represents the position of the phosphate with respect to the complex, while the pink sphere denotes the position of the hydrogen bond between bases. The blue sphere approximates the excluded volume of the nucleotide. \textbf{(b)} This panel shows the main interaction sites between consecutive beads in the same strand. The equilibrium distance between patches (E-F) is set to 0.34 nm while the one between beads centres (A-B) to 0.46 nm. This leads to an equilibrium distance of 0.7 nm between the external edge of the backbone (C-D). These distances are set so that the correct pitch of $10$ bp is recovered. \textbf{(c)} Two nucleotides are bonded via a breakable harmonic spring. Their distance is set so that the full chain thickness is around $2$ nm, as that of B-DNA. \textbf{(d)} Representation of the double-stranded DNA model. The red chain also shows the beads which interact sterically (solid red) as well as the phantom beads (solid grey). The faded red spheres represent the steric interaction volume of the red beads. Neither the interacting beads nor the ghost beads along the blue chain are shown to ease visualisation.
	\label{model}                 
}
\end{figure}

\section{The model}
\label{sec:Model}

We start by considering a complex made of two spherical monomers (see Fig.\ref{model}(a)), one of which represents the sugar-phosphate backbone (``bead'' hereafter, shown in blue), while the other represents the nitrogenous base (``patch'' hereafter, shown in violet), and is placed at a distance of 0.5~nm from the bead centre. Beads have an excluded volume so that they cannot overlap, whereas patches have no associated excluded volume. In order to see the resolution of the system, a fictitious nucleotide structure lying inside the bead is shown in black in Fig.~\ref{model}(a). Although a phosphate group is not directly included in the current version of the model, is  marked in yellow in Fig.~\ref{model}(a) for clarity; this is sitting 0.5~nm from the bead centre but slightly away from the antipodal point to the patch. Each bead-patch complex represents a single nucleotide, and acts as a rigid body; we connect a chain of these bodies via FENE bonds of length $d_{bp}=0.46$~nm between the beads to represent one strand of DNA. We set the distance between two consecutive patches along the strand (E-F in Fig.~\ref{model}(b)) at $0.34$ nm by means of a Morse potential; the difference between the lengths A-B and E-F implies that the distance between the implicit phosphates at the external edge of the beads (C-D in Fig.~\ref{model}(b)) is $d_{ph}=0.7$ nm. The ratio between $d_{bp}$ and $d_{ph}$ is well known to crucially regulate the correct pitch of the chain~\cite{Calladine1997} (for details about the potentials used see Appendix~\ref{apendixa}).

Nucleotides belonging to different strands are bonded together with breakable harmonic springs between two patches, representing hydrogen bonds (see Fig.\ref{model}(c)). The equilibrium bond distance is set to zero; if the extent of the bond increases beyond a critical value $r_{c}=0.3$~nm, the bond breaks, modelling denaturation. 

While the pitch of the chain is set by the ratio of the base pairing distance and the distance between successive phosphate groups on a DNA strand, the right-handedness is imposed using a dihedral potential between the quadruplets of monomers forming two consecutive nucleotides (A, E, F and B in Fig.\ref{model}(b)). This potential regulates the angle between the planes A-E-F and E-F-B. The minimum of this potential is arbitrarily set at 36$\degree$, as this matches the geometry of a regular dsDNA helix. 

In order to limit the splay of consecutive nucleotides (also called ``roll''~\cite{Calladine1997}) we used a stiff harmonic potential so as to keep the angle between particles E, F and B (two patches and one bead) at $90\degree$ (Fig.~\ref{model}(b)). This interaction imposes the planarity between consecutive bases in the same strand. Finally, the last ingredient of this model is a Kratky-Porod potential regulating the angle between three consecutive patches along one strand. This allows us to finely regulate the chain stiffness.

The excluded volume around each bead depicted in Fig.~\ref{model}(d) (faded red spheres) has diameter $1$~nm. Since we use spherical beads rather than asymmetrically shaped ones (this is important for the speed of the algorithm), the geometry of the double-strand depicted in Fig.~\ref{model}(b) and (d) would involve a large degree of overlapping which would lead to a large steric repulsion. To avoid this we consider two types of beads in each strand: sterically interacting beads (shown as small solid red spheres for one strand in Fig.~\ref{model}(d)) are intercalated by two ghost beads (depicted as small grey spheres) which don't interact sterically along the same strand but they do interact with all the beads on the complementary strand with an excluded volume of 0.5~nm. This choice  ensures that only non-overlapping beads sterically interact with one another. In addition, this allows us to preserve the correct thickness of the chain ($2$ nm for B-DNA), to maintain the desired distance between contiguous nucleotides and avoid the strands crossing through one another. 

This model is based on few crucial geometric constraints of double-stranded DNA while the aim of delivering large-scale simulations is achieved by using spherical monomers that interact via standard potentials. These are efficiently implemented in LAMMPS and deliver a highly scalable performance in large scale parallel simulations (see Appendix~\ref{apendixa} for more details). In Fig.~\ref{equilibrated} we show a typical equilibrated configuration using the presented model for a 1000~bp molecule.  
 

\begin{figure}[H]    
	\begin{center}
		\includegraphics[width=0.4\textwidth]{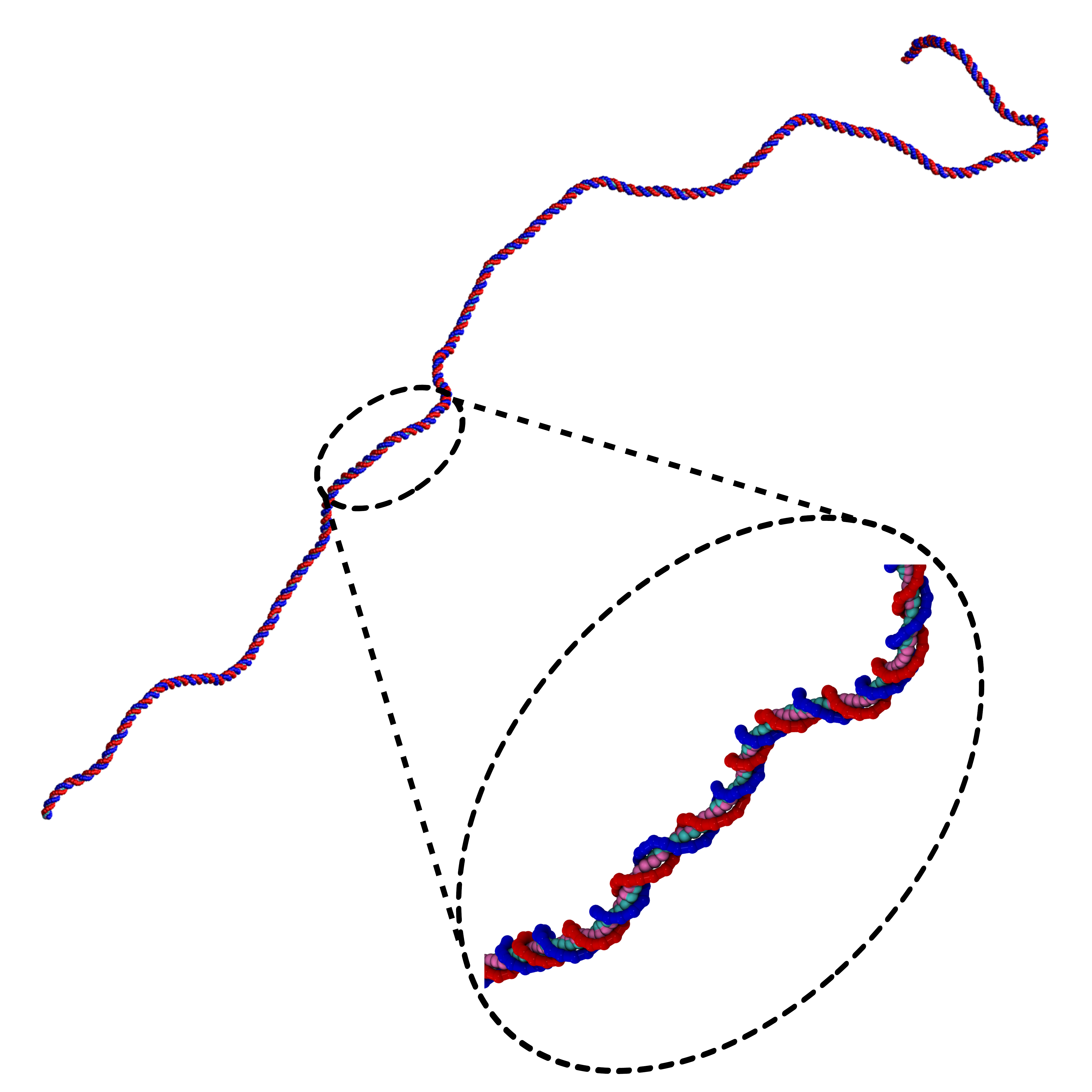}
	\end{center}
	\caption{An example of an equilibrated configuration of a 1000 bp double-stranded DNA molecule, as simulated with the model presented in Sec.~\ref{model}.}
	\label{equilibrated}                 
\end{figure}

\section{Parameterisation}
\label{sec:Validation}

Our model has several parameters which can be varied to control the pitch, bending and torsional properties of the simulated DNA molecule. Nonetheless, we are interested in modelling the B form of dsDNA, of which two main physical properties are: the persistence length $l_p = 50$ nm $\simeq 150$ bp, and the torsional rigidity $C/k_BT \simeq 60 - 80$ nm $\simeq 177 - 235$ bp~\cite{Lipfert2010,Bryant2003a,Strick1999}. Due to the interplay between the potentials presented in the previous Section, there is no simple mapping between individual simulation parameters and the resulting physical properties; instead we obtain a simulated molecule with the correct values of $l_p$ and $C$ via a systematic tuning of the parameters. In this Section we measure these properties from the microscopic positions of the beads in equilibrated DNA molecule configurations. Then in the following Section, we use the parametrised force field to simulate single-molecule experiments, showing that the DNA molecules show the correct macroscopic response to mechanical manipulations.

\subsection{Persistence Length}

\begin{figure}[t]
\begin{center}
\includegraphics[width=0.5\textwidth]{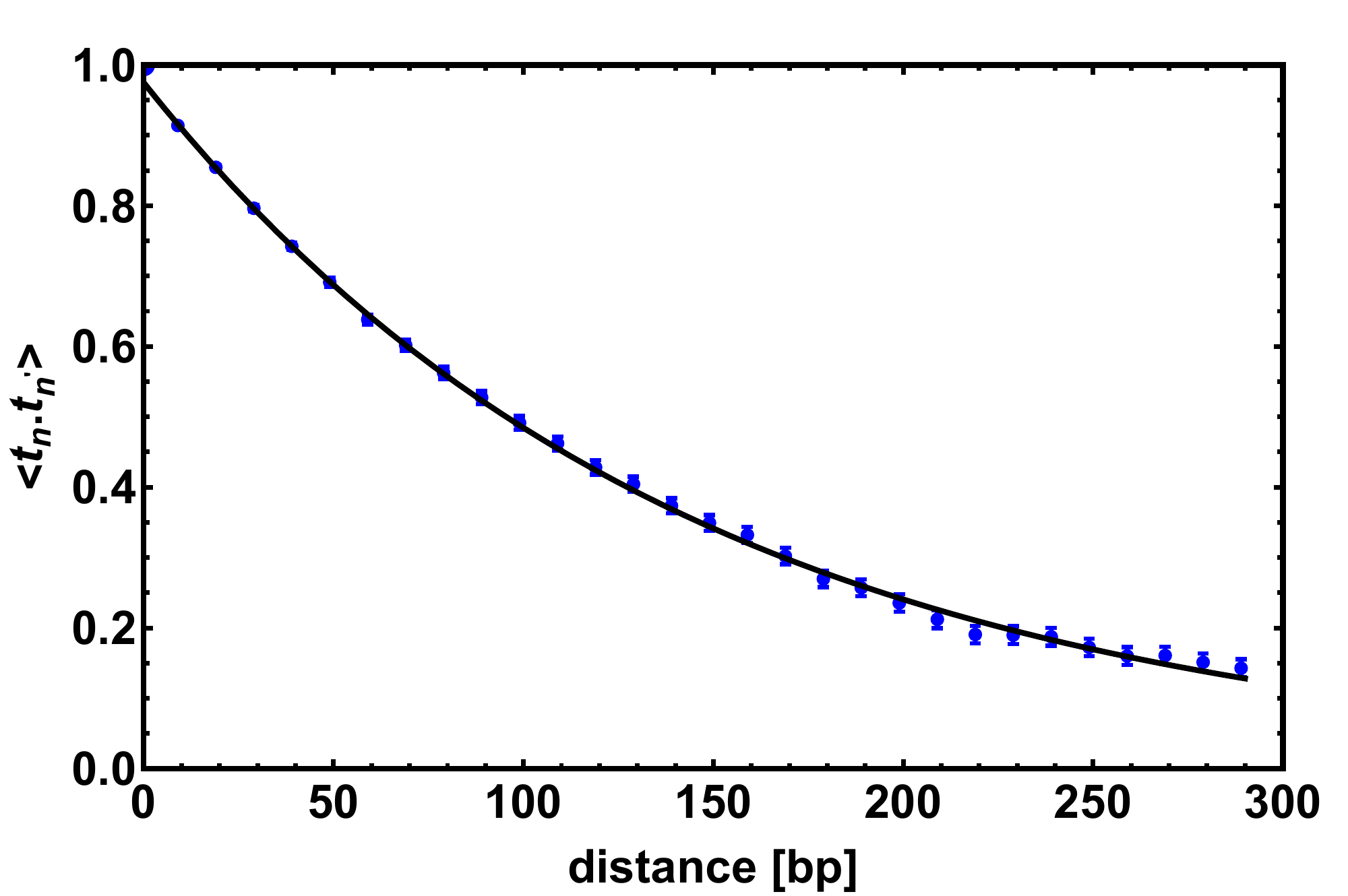}
\end{center}
\vspace*{-0.5 cm}
\caption{The tangent-tangent correlator $\langle {\bf t}(n) \cdot {\bf t}(n^\prime)\rangle$ computed for a chain 300 bp long; it shows an exponential decay as in Eq.~\eqref{tangentcorrelation} with a decorrelation length $l_p = 143 \pm 7$ bp. Points show correlations measured from the simulations (average over time), and the line shows a fit to Eq.~\ref{tangentcorrelation}. Error bars give the standard error in the mean. \label{persistence}}
\end{figure}

The persistence length of dsDNA is a well-studied physical property that plays an important role in the wrapping of dsDNA around histone octamers to form the chromatin fibre, as well as in many other biological processes. In physical terms it gives a measure of the length-scale over which the direction of the chain is no longer correlated with itself.
Following the description of an elastic rod by Moroz and Nelson~\cite{Moroz1998} one can define the bending rigidity via the elastic energy functional
\begin{equation}
	\dfrac{E_{\rm bend}}{k_BT} = \dfrac{l_p}{2} \int_0^L \left( \dfrac{d{\bf t}}{ds}\right)^2 ds 
\end{equation}
where $l_p$ is the bending persistence length, $s$ the arclength parameter and $\mathbf{t}(s)=d\mathbf{r}/ds$ the tangent to the chain (at $s$) whose location in space is described by $\mathbf{r}(s)$. This quantity can also be readily measured by computing the tangent-tangent correlator:
\begin{equation}
\langle {\bf t}(s) \cdot {\bf t} (s^{\prime}) \rangle = \displaystyle e^{ -\lvert s-s^{\prime} \rvert /l_{p}}.
\label{tangentcorrelation}
\end{equation}
In our model, we use the position of the patches to extract the centreline of the dsDNA molecule, where the tangent at the $n$th patch at position ${\bf r}(n)$ is ${\bf t}(n) \equiv ({\bf r}(n+1)-{\bf r}(n))/\lvert {\bf r}(n+1)-{\bf r}(n) \rvert$. One can compute the tangent-tangent correlator along this curve and  obtain the persistence length by extracting the exponent of the exponential decay. In order to avoid finite-size effects due to the presence of ends, we neglect the two terminal segments ($\sim$5 bp at each end). The resulting curve is shown in Fig.~\ref{persistence}. The exponential fit returns a persistence length $l_{p} \simeq 143 \pm 7$ bp, in agreement with experimentally observed values.

\subsection{Torsional Rigidity}\label{tors_sec}

The behaviour of DNA when twisted is regulated by its torsional rigidity. There are several well known examples in which this property is crucial for important biological processes, such as transcription and gene expression~\cite{Ding2014,Brackley_supercoil}. Furthermore, the high torsional stiffness of DNA molecules implies that, when placed under torsion, they preferentially bend, thereby creating writhe and plectonemes~\cite{Calladine1997}. In order to take this feature correctly into account, it is therefore crucial to accurately model the competition between bending and torsional rigidities~\cite{Strick1999}. 

Following Moroz and Nelson~\cite{Moroz1998} once again, we first define the torsional stiffness of an elastic rod $C$ via the elastic energy functional
\begin{equation}\label{tors_eq1}
	\dfrac{E_{\rm tors}}{k_BT} = \dfrac{C}{2} \int_0^L \Omega_3(s)^2 ds,
\end{equation}
where $\Omega_3(s)$ is the rate of rotation of a local reference frame along the curve around the tangent ${\bf t}(s)$, defined as in the previous section.  

\begin{figure}[t]    
\begin{center}       
\includegraphics[width=0.5\textwidth]{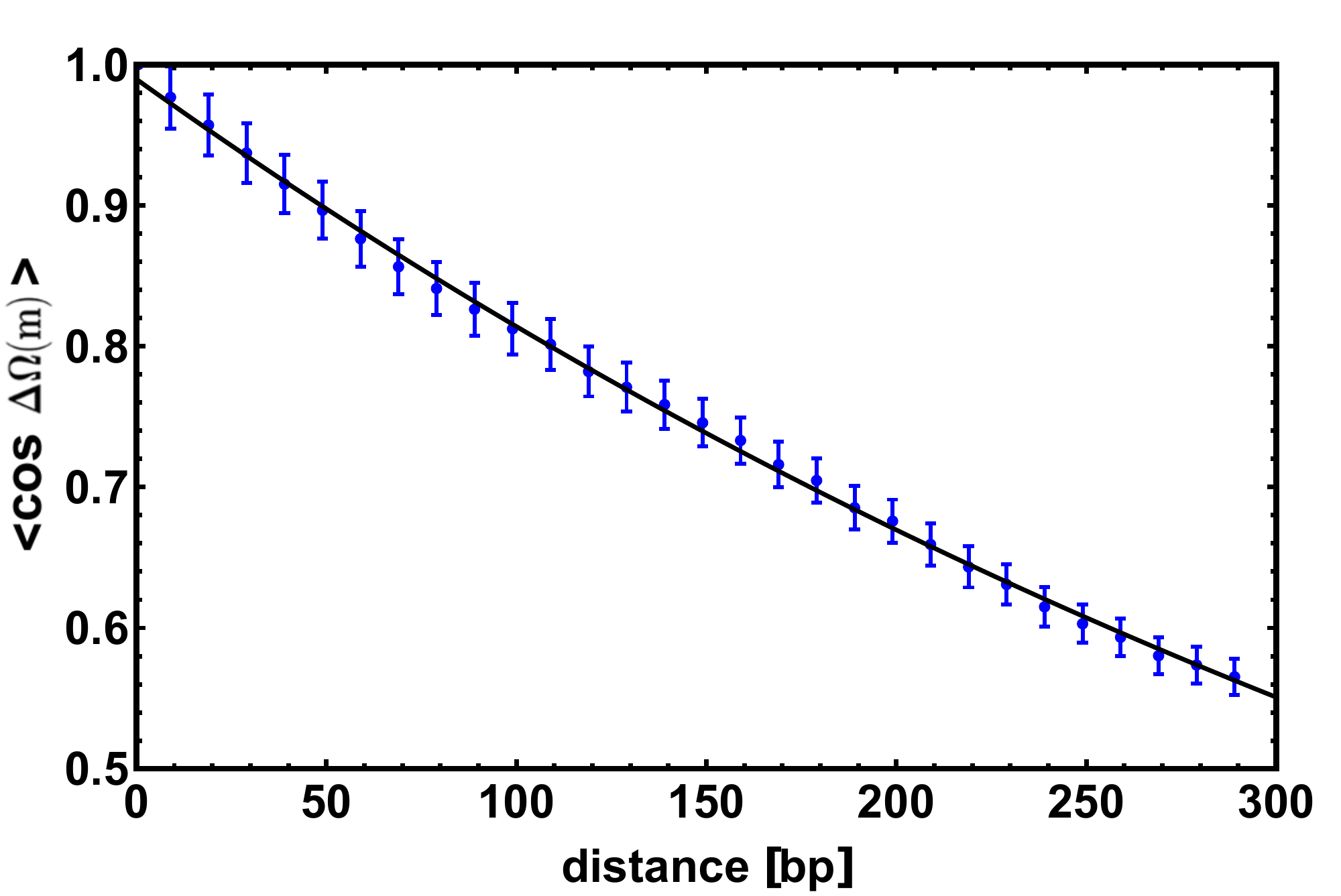}
\end{center}
\vspace*{-0.3 cm}
\caption{The average of the cosine of the total twist angle $\Delta\Omega(m)$  is computed for a chain 300 bp long; in this figure we show the correlator to decay exponentially as in Eq.~\eqref{torsionalpersistence} with a characteristic length $l_\tau = 512 \pm 18$ bp. Data points are obtained from simulations while the line is an exponential fit with $f(d)=e^{-d/l_\tau}$.
\label{torsionalpl}}
\vspace*{-0.3 cm}
\end{figure}

Analogous to the measurement for the bending persistence length via the tangent-tangent correlator, we here measure the torsional persistence length by computing the decorrelation of the twist angle. This correlator can be quantified by defining a local reference frame for each base pair, and tracking the transformation of the frames from one base pair to the next via the Euler angles. Each local frame is specified by the tangent vector ${\bf t}(n)$ as defined above, a normal vector ${\bf f}(n)$, defined as the projection of the vector connecting two beads in a base-pair, onto the plane perpendicular to ${\bf t}(n)$, and a third vector vector ${\bf v}(n)={\bf t}(n) \times {\bf f}(n)$, perpendicular to both ${\bf t}(n)$ and ${\bf f}(n)$.

The Euler angles between the frames at $n$ and $n+1$ can be used to obtain the twist increment between those base-pairs, and the correlation function for the total twist between $m$ consecutive base-pairs $\Omega(m)$ calculated. Since the DNA has an equilibrium twist angle $\theta_0=36\degree$ per bp, we subtract this out, and calculate the correlation for the residual twist $\Delta\Omega(m)=\Omega(m)-m\theta_0$. 
It can be in fact shown~\cite{Brackley2014a} (see also Appendix~\ref{ap_tors}) that the average cosine of the residual total twist between any two reference frames separated by $m$ bases exhibits an exponential decay as:
\begin{equation}
\langle \cos{\Delta\Omega(m)} \rangle = \displaystyle e^{-m / 2C}, 
\label{torsionalpersistence}
\end{equation}
where we define $l_{\tau}=2C$ the characteristic torsional correlation length. We obtained $\langle \cos{\Delta\Omega(m)} \rangle$ from a 300 bp long DNA molecule and averaged it over time. The curve obtained is shown in Fig.~\ref{torsionalpl} on top of which we show the fitted exponential which has a characteristic decay length $l_{\tau} = 512 \pm 18 \text{ bp} \simeq 174 \pm 6 \text{ nm}$, which is consistent with experimental estimates valid for the B-form of dsDNA. 

\section{Validation through single molecule experiments}
\label{sec:SingleMolExp}

Many cellular processes, such as replication and transcription, are carried out by proteins acting on single DNA segments. In light of this, recent years have seen an increasing interest in experimental techniques such as optical tweezers and atomic force microscopy, that can probe the response of DNA to external stresses (modelling the effect of DNA-binding enzymes) at the single-molecule level. In particular, the stretching and twisting behaviour of DNA under external forces and torques has been thoroughly  investigated~\cite{Strick1999,bustamante10years,bustamantesmallforce,stretchingwithoptical,Bryant2003a,Ritort2006a,Lipfert2010}.

In this section we aim at reproducing the conditions of two different experiments, in order to test the response of our model DNA to stretching and twisting. This also provides us with an independent method to evaluate its persistence length and torsional rigidity. In the following, we therefore keep the parameters fixed at the values used in the previous Section, and do not further tune them to achieve the experimentally known behaviours.

\subsection{Response to Stretching}

\begin{figure}[t]    
	\begin{center}       
		\includegraphics[width=0.48\textwidth]{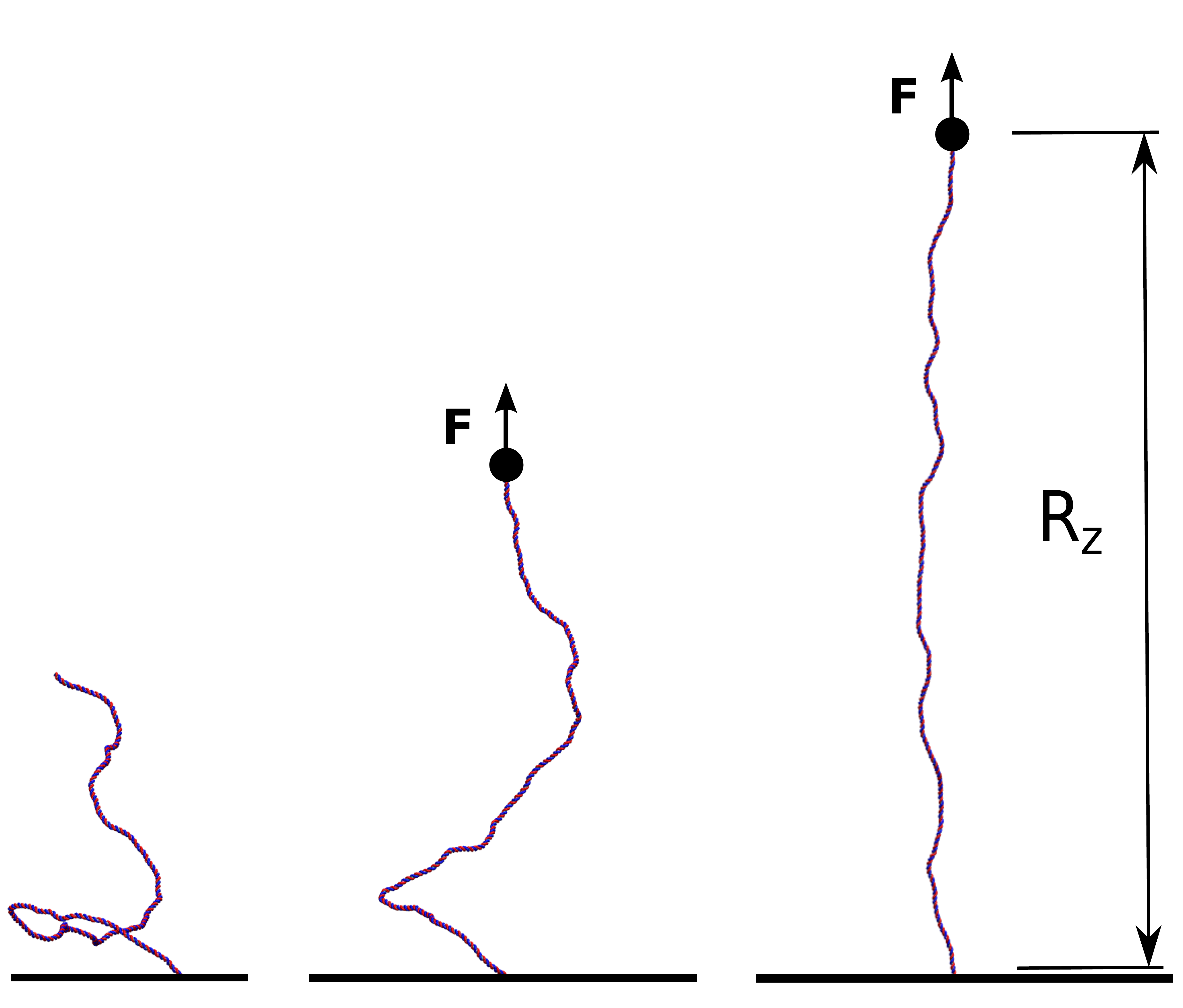}
	\end{center}
	\vspace*{-0.3 cm}
	\caption{In order to simulate single-molecule experiments the model for dsDNA is anchored to a surface at the bottom end while being stretched with a constant force {\bf F} from the top end. We then monitor the end-to-end elongation along the $z$-direction, $R_{z}$, and report its equilibrium value for a given force in Fig.~\ref{fve}.}
	\label{stretch_exp}                
	\vspace*{-0.3 cm}
\end{figure}

The classic elastic response of DNA to an external stretching force ${\bf F}$ is that of an entropic spring with relaxed length $R_0 \sim N^\nu$ with $\nu=0.588$ for a self-avoiding polymer. The force required to induce an end-to-end distance $R_{z}= [{\bf r}(L)- {\bf r}(0)] \cdot {\bf e}_{z}$ for a chain of length $L$ and persistence length $l_p$ can be approximated using the worm-like chain (WLC) result~\cite{smithsmallforce,marko-siggia}: 
\begin{equation}
\frac{F l_{p}}{k_{B} T} = \frac{R_{z}}{L} + \frac{1}{4(1- \frac{R_{z}}{L})^{2}} - \frac{1}{4},
\label{stretchWLC}
\end{equation}
where excluded volume effects are neglected (a good approximation when $L$ is not much larger than $l_p$, as in our case).
In order to test this result we performed simulations in which a constant pulling force directed along ${\bf e}_z$ and acting on the last base pair of the dsDNA was applied, while the other end of the molecule was anchored at a surface (see Fig.~\ref{stretch_exp}). 

\begin{figure}[t]    
	\begin{center}       
		\includegraphics[width=0.5\textwidth]{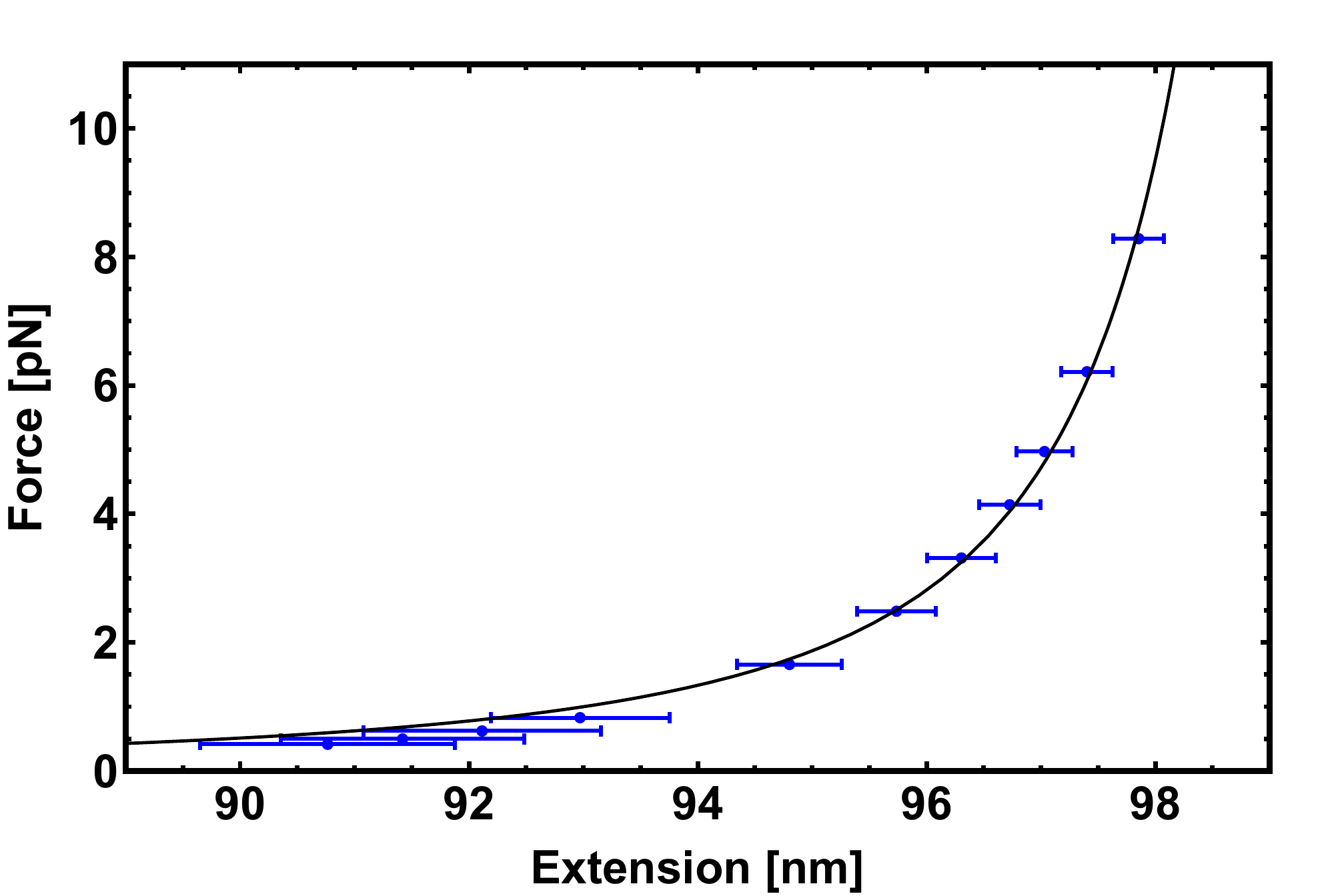}
	\end{center}
	\vspace*{-0.5 cm}
	\caption{Force-extension curve from the simulation (data points) and fitted by the function in Eq.~\eqref{stretchWLC} (solid line). The free parameters for the fitting are the total polymer length $L$ and the persistence length $l_p$, both of which are in agreement with the fixed parameters of the model (see text). }
	\label{fve}
	\vspace*{-0.5 cm}
\end{figure}

The force-extension curve~\cite{Ritort2006a} observed for a chain 300 bp long 
is reported in Fig.~\ref{fve} as data points, while the solid curve is the fit to Eq.~\eqref{stretchWLC}. The fitting results in values for both $L$ and $l_p$, that we can compare with the values set in our model. In particular for a 300~bp chain we obtain $L=100.3 \pm 1.7$~nm (which gives a bp step size of $0.33 \pm 0.01$~nm) and $l_p=47 \pm 2$ nm $\simeq 140 \pm 7$ bp. When $l_p$ is measured from the tangent-tangent correlation for the same chain without applied force a value of $l_p=49$~nm was obtained. The results are therefore in good agreement with the calculation and the tuning of the persistence length performed in the previous Section.


\begin{figure}[t]    
	\begin{center}       
		\includegraphics[width=0.33\textwidth]{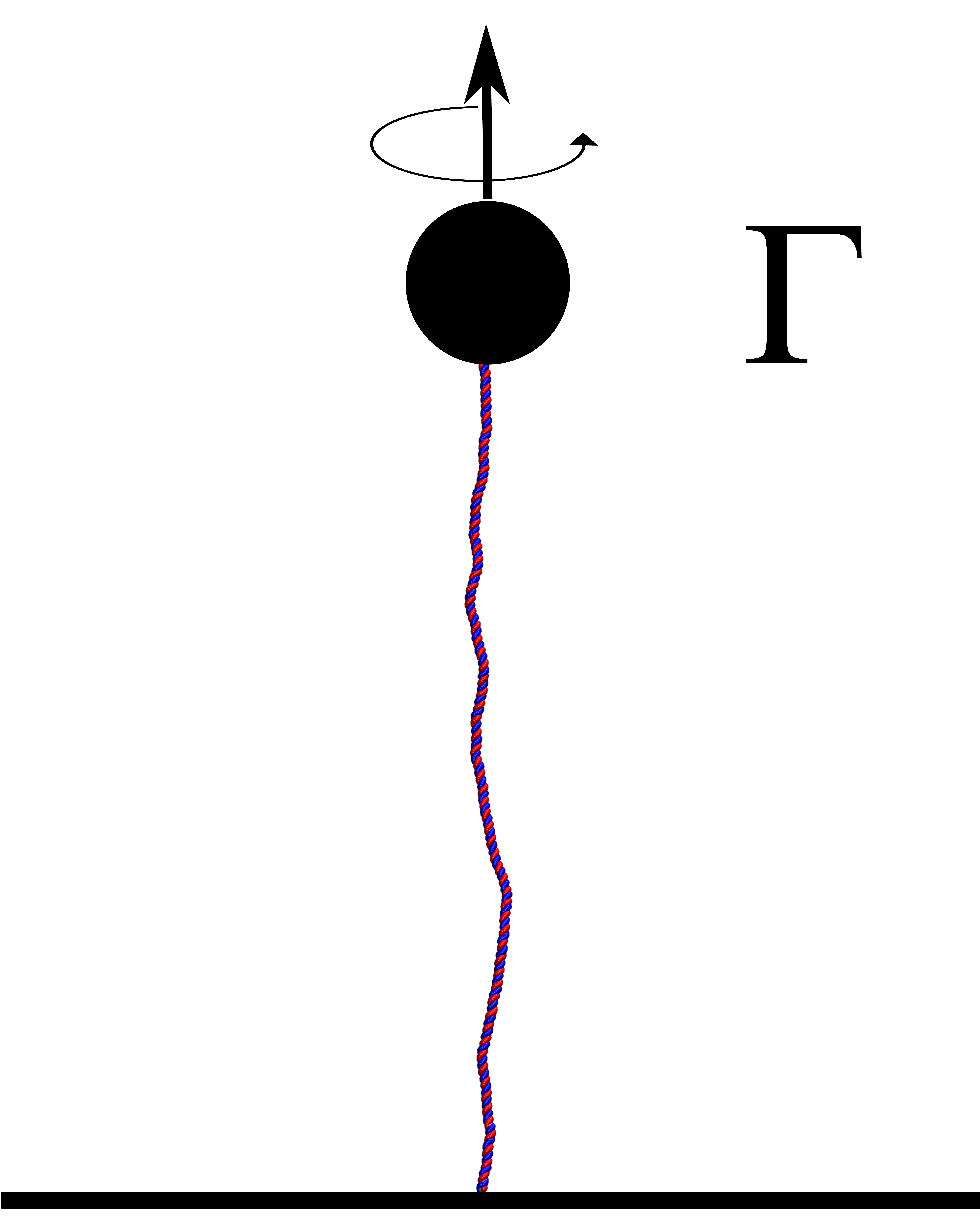}
	\end{center}
	\caption{The model DNA is anchored to a surface at the bottom end while being stretched with a constant force {\bf F}, and a torque $\mathbf{\Gamma}$ is applied at the top end. We then monitor the linking number and report its equilibrium value for a given torque. With this information is possible to compute the superhelical density.}
	\label{torque_exp}                 
\end{figure}

\begin{figure}[t]    
	\begin{center}       
		\includegraphics[width=0.5\textwidth]{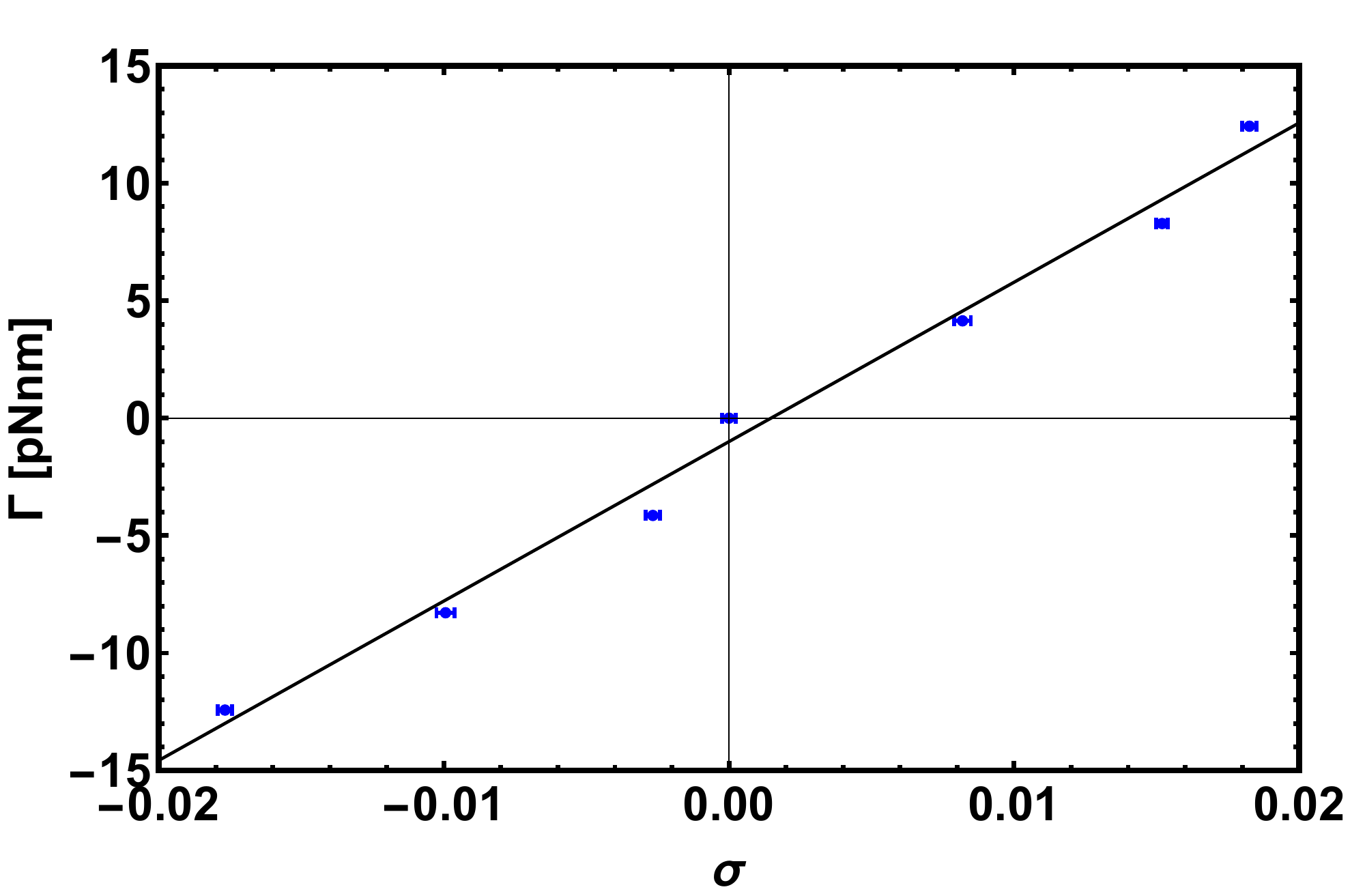}
	\end{center}
	\caption{Response to torque experiment. Here we show the linear regime for small $|\sigma|$. Fitting the data points gives the torsional rigidity $C$ using Eq.~\eqref{eq:tors_rigidity}. }
	\label{tvsh}                 
\end{figure}

\subsection{Response to Twisting}

The torsional stiffness of DNA can be calculated by computing the twist response of dsDNA to an imposed external torque, for instance applied by a magnetically controlled macroscopic bead~\cite{Matek,Strick1999} (see Fig.~\ref{torque_exp}). For different magnitudes of the applied torque, $|\bm{\Gamma}|$, we compute the superhelical density, $\sigma$. The level of supercoiling is determined by the linking number $Lk$, the number of times one DNA strand wraps round the other. 

Since a dsDNA chain has a preferred equilibrium linking number $Lk_0$, the superhelical density is defined as $\sigma=(Lk-Lk_0)/Lk_0$. The well-known White-Fuller theorem~\cite{Fuller1978} 
\begin{equation}
 Lk = Tw + Wr.
  \label{eq:linking}
\end{equation}
relates the linking number of the edges of  a ribbon ($Lk$) to the twist ($Tw$), \emph{i.e.} the extent of rotation of the two ribbon edges about the axis, and the writhe ($Wr$), \emph{i.e.} the self-crossing of the ribbon centreline. Although the chain we use is not closed into a loop, and therefore it is not possible to formally define a linking between the strands, it is possible to compute the linking number between two ``artificially'' closed strands~\cite{Orlandini2000,Michieletto2014e} which follow the paths of the DNA strands along the chain backbone and then join the respective ends far away from the molecule (see Appendix \ref{closure_lk}). By applying a force to the molecule, we keep it straight, consequently imposing null writhing and, in turn, ensuring that the twist is equal to the computed linking number. 
By measuring the deviation of twist $\Delta Tw$ from the equilibrium value $Tw_0$ defined as the number of base-pairs divided by the pitch $p=10$ bp, we can readily obtain $\sigma$. 


 With this information it is possible to recover the response curve of the molecule to an external torque. A feature of this is a linear regime for small $\lvert \sigma \rvert$ which we recover (see Fig.~\ref{tvsh}). The torsional rigidity, $C$, can finally be calculated as~\cite{Moroz1998}
\begin{equation}
C=\dfrac{1}{k_BT}\dfrac{a_{0}}{\theta_{0}} \dfrac{\Delta \Gamma}{\Delta \sigma},
\label{eq:tors_rigidity}
\end{equation}
where $a_{0}=0.34$ nm is the double helical rise for a relaxed dsDNA and $\theta_{0}$ is the equilibrium twist angle across a base-pair step in the relaxed case.
The data points shown in Fig.~\ref{tvsh} are obtained from simulations of a 600 bp long chain anchored at a surface to one end, while the other end was pulled by a constant force of $16$ pN and different torques, $\Gamma = {\bf \Gamma}\cdot {\bf e}_z$, were applied. From the fit we get the value of torsional persistence length $C = 88 $ nm $\simeq 260$ bp in good agreement with experimental results~\cite{VologodskiiBook,Bouchiat1999,Oroszi2006}.
One can finally use the relation between the torsional persistence length $l_\tau$  and the torsional stiffness $C$ obtained from the twistable worm-like chain theory~\cite{Brackley2014a}, which gives $l_{\tau}=2C = 176$ nm, very close to the measurement performed in the previous section ($l_{\tau}=174 \pm 6$ nm).

\section{DNA Denaturation and Supercoiling}

DNA denaturation is the separation and unwinding of the two strands,  transforming a DNA duplex into two isolated and unbound single strands~\cite{Poland1966a}. This process can be driven by heating a solution of dsDNA molecules, and a critical ``melting'' temperature $T_m$ can be defined as the temperature at which 50\% of a long dsDNA molecule is denatured. This critical temperature commonly depends on the genetic sequence, pH and salt concentration~\cite{Kirk1967,Gruenwedel1969,VologodskiiBook}. 
Localised, temporary, and dynamic denatured segments are often referred to as ``bubbles''. 

It is well known that local denaturation has several biological implications such as favouring transcription initiation, DNA repair or recombination~\cite{Botchan1973,Kabakcioglu2009a,Ding2014} and that the dynamics of these bubbles can be affected by torsional stress, which is itself often regulated by enzymes, such as RNA polymerases~\cite{Lia2003,Jeon2010,Lavelle2014}. This fascinating interplay between the elasticity and biology of DNA has received much theoretical and experimental attention~\cite{Benham1974,Benham1979,Hatfield2002,Carlon2002,Kabakcioglu2009a,Lavelle2014,VologodskiiBook}, but there have been remarkably few attempts to tackle it from a computational point of view~\cite{Mielke2005,Sicard2015}. Whereas theoretical models can capture the thermodynamics of a ``stress-induced DNA-duplex destabilisation'' (SIDD)~\cite{Wang2008}, elucidating the kinetics of such a process, under both equilibrium and out-of-equilibrium conditions, is an important question that can be addressed using numerical investigations. 

In this Section we show that our model can readily recapitulate DNA denaturation upon decreasing the stiffness, $K_{2}$, of the spring connecting patches in the two strands ($U_{hb}$). While the most common strategy is increasing the solution temperature, here we focus on a pathway that more closely mimics a change in salt concentration~\cite{Gruenwedel1969} or solution pH. 

\begin{figure}[t]    
	\begin{center}       
		\includegraphics[width=0.5\textwidth]{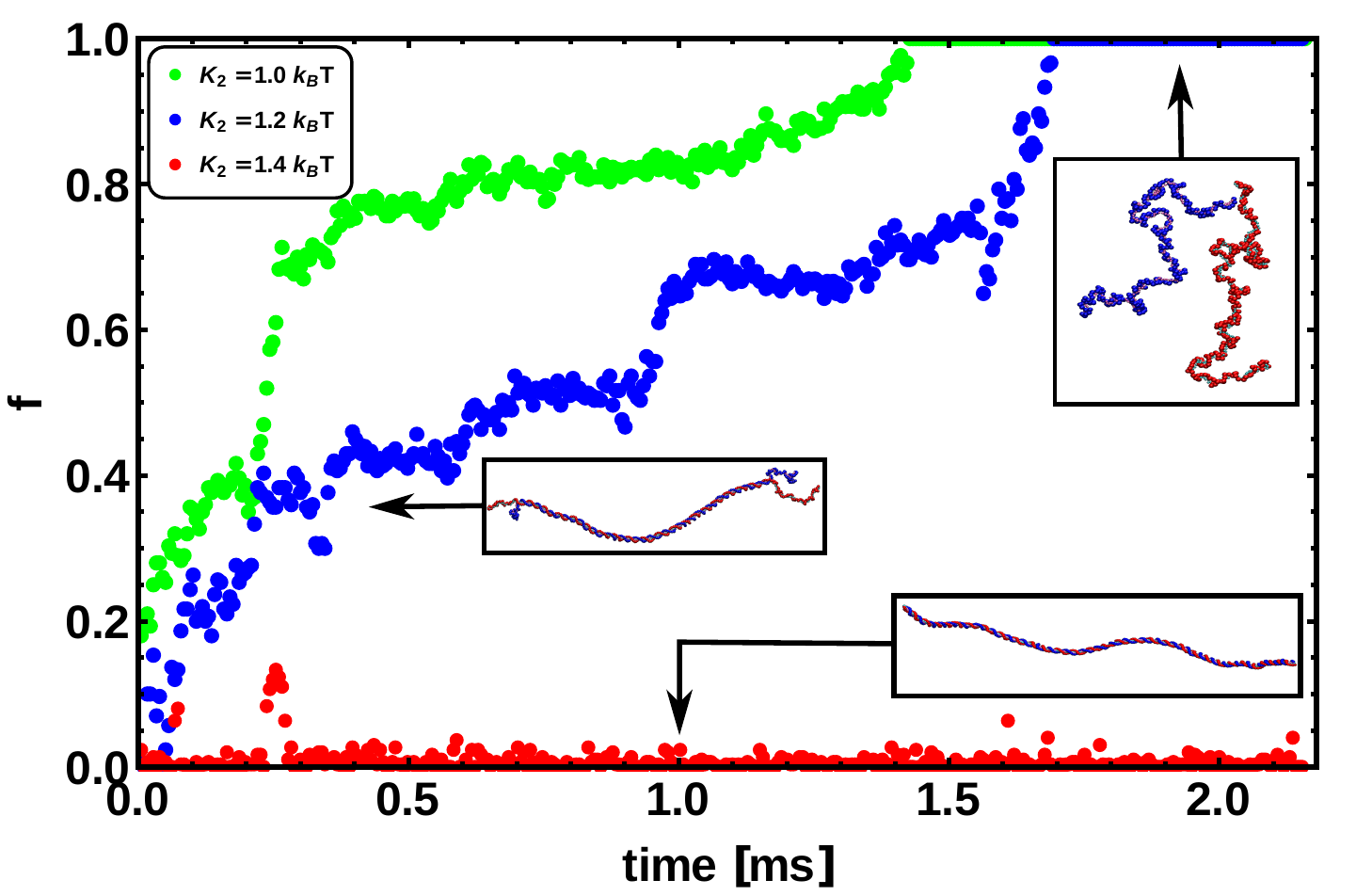}
	\end{center}
	\vspace*{-0.5 cm}
	\caption{This figure shows the fraction of denatured base pairs $f$ as a function of time and for different bond energies connecting the patches of paired bases. Snapshots from simulations are also shown. The energies used range between $K_{2}=1.0~k_{B}T$ and $K_{2}=1.4~k_{B}T$. We always observe that in linear dsDNA the denaturation process nucleates from the ends, as argued from experiments~\cite{Beers1967}. }
	\label{den3}                
	\vspace*{-0.5 cm}
\end{figure}

In Fig.~\ref{den3} we show the fraction of denatured base-pairs as a function of time for three different choices of $K_2$. As the energy of the bond is decreased, we observe the unbinding of two strands nucleating from the ends of the chain, as observed experimentally~\cite{Beers1967}. We then observed that the denaturation spreads to the middle of the molecule, finally melting the  whole chain when $K_2 \lesssim 1.2 k_BT$ and producing two single strands.

Single stranded DNA (ssDNA) is much more flexible than its bound counterpart. In order to mimic this behaviour in our model, we eliminate both the dihedral and the Kratky-Porod interactions between nucleotides which are part of a ``bubble'' larger than two base-pairs. This results in single strands with a persistence length of around 2 bp which are extremely flexible, as one can appreciate from the snapshots in Fig.~\ref{den3}.

It is worth stressing that while our model can show the reverse of partial denaturation, by for instance increasing $K_{2}$ back to higher values, it cannot create hybridised molecules in which bases pair with partners other than those which they started with (i.e. secondary structures cannot form). This is a limit of the current model which we aim to improve in the future.

It is also worth highlighting that this single-nucleotide resolution model can, in principle, readily incorporate sequence specificity. This can be done, for instance, by defining two types of harmonic bonds connecting patches in the complementary strands and by using springs with different stiffness such that $K_2(AT) < K_2(CG)$. In light of this, we expect that this model, thanks to its high scalability when run in parallel, will be of use to investigate the dynamics of denaturation in long dsDNA molecules, whether torsionally relaxed or supercoiled. 

\begin{figure}[H]    
    \begin{center}       
        \includegraphics[width=0.5\textwidth]{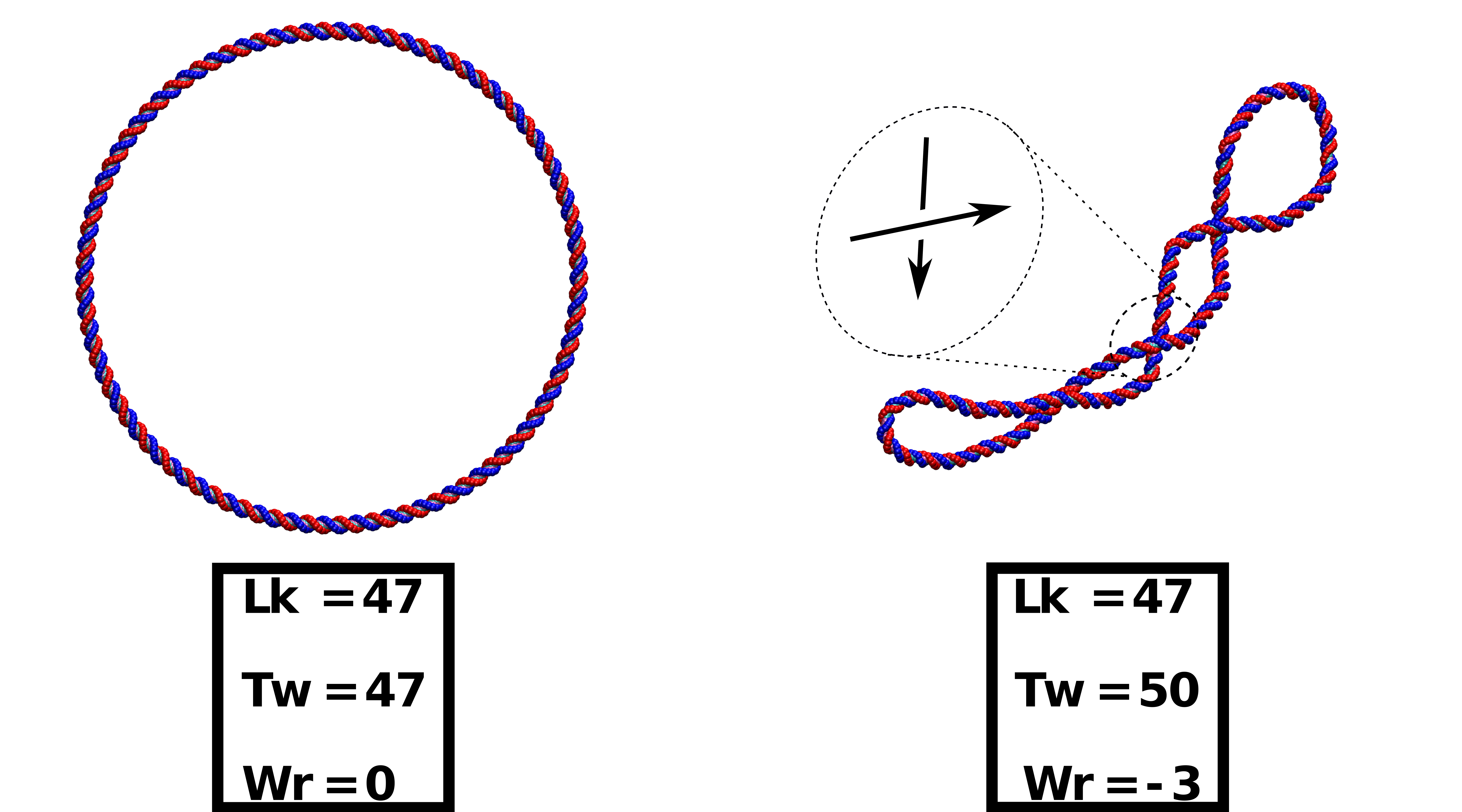}
    \end{center}
\caption{This figure shows the relaxation of a negatively supercoiled circular dsDNA. (left) The molecule (500 bp long) is initialised as a perfect ring from which three full turns are removed. (right) As the system evolves, the lack in twist is converted into writhe, and the molecule assumes stable buckled configurations. This behaviour is expected for a real dsDNA molecule because the torsional stiffness is larger than the bending rigidity.}
    \label{sc}                 
\end{figure}

As a preliminary step to show that our model can readily take into account supercoiling, in Fig.~\ref{sc} we give an example of a simulation for supercoiled DNA. A model dsDNA ring of contour length equal to 500 bp is initialised with a linking number deficit of $\Delta Lk=Lk_0-Lk=-3$ (47 turns instead of the usual 50 for a pitch of 10 bp). In a linear molecule this deficit would be quickly washed out by the free motion of the ends, whereas in a closed molecule, the difference creates a negative supercoiling $\sigma = \Delta Lk/Lk_0 \simeq -0.06$ which is conserved throughout the dynamics. The amount of supercoiling can then be distributed into the torsional or bending degrees of freedom as long as the White-Fuller theorem~\cite{Fuller1978} is satisfied  (see Eq.~\eqref{eq:linking}). Since the torsional stiffness of DNA is bigger than the bending rigidity, much of the twist is quickly converted to writhe, as can be readily seen in Fig.~\ref{sc}.

\section{Discussion}
\label{sec:Discussion}
The interplay between the physics and biology of DNA is one of the most intriguing topics in biophysics. While computational models can strongly aid the understanding of this fascinating open problem, the computational resources for such an expensive task have traditionally been limited. Researchers often use either very detailed and accurate all-atoms models, which can only cover short time and length scales, or coarse-grained models, which can follow the evolution of the system for much longer times, but at the expense of neglecting key physical properties of dsDNA. Mesoscopic models have been recently proposed to fill in the gap between these two approaches~\cite{Ouldridge2011a}, but they have not yet been exported to a highly efficient and parallel environment. Here, we have proposed a mesoscopic model that can be readily implemented at minimal cost into LAMMPS, one of the most popular molecular dynamics codes for atomistic and mesoscopic simulation. 

Our model aims at bridging the gap between all-atoms and coarse-grained models for dsDNA; while it is currently less sophisticated than other mesoscopic models, most notably in the treatment of sequence specificity or hybridisation, this model exploits the scalability of LAMMPS, and is ideally suited to study problems such as DNA-protein interactions, or the denaturation of supercoiled DNA, where it is essential to consider long molecules, as well as to simultaneously model double-stranded and denatured regions.

This model can also be extended to include base-pair specificity, and variable salt or pH concentration, while allowing the user to reach biologically relevant time and length scales. In this paper we have shown that this model is capable of reproducing DNA melting and, more importantly, of tracking the dynamics of supercoiled molecules $\sim 1000$ bp long for up to $\sim 2$ ms. In the near future, we aim to use this model to investigate further the interplay between denaturation and supercoiling, especially in light of its connection to gene expression~\cite{Brackley_supercoil,Ding2014}.

We should also highlight that the presented model has several limitations which arise from the compromise between accuracy and scalability. For instance, our model lacks the ability of reproducing realistic hybridisation events where distant parts of the chains can become bonded forming an intermediate hairpin. The choice of neglecting the modelling of such events allows us to employ short-ranged neighbour calculations -- \emph{i.e.} the algorithm does not include $O(N^2)$ loops -- which markedly improves the computational speed-up. 

\begin{figure}[t]    
	\begin{center}
		\includegraphics[width=0.5\textwidth]{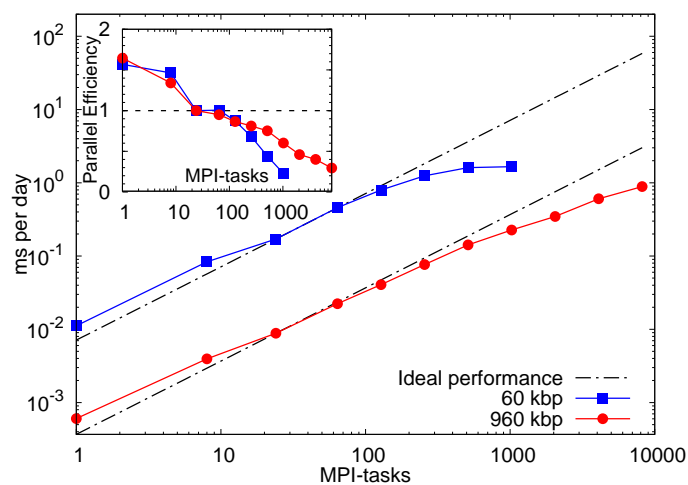}
	\end{center}
	\caption{This plot analyses the strong scaling behaviour of the model. The figure shows the simulation time in microseconds per day as a number of processes (MPI-tasks). Two different benchmarks were used, a small one with $60$ kbp and a 16 times larger one with $960$ kbp. The results are compared with the timings of a run with 24 processes for each benchmark, corresponding to one fully occupied node on the ARCHER XC30 architecture. This leads to parallel efficiencies (see inset) in excess of $100\%$ for $1$ and $8$ processes. Total simulation times of up to $2$ ms per day are feasible.}
	\label{scale}                 
\end{figure}

Furthermore, we also extensively tested the scalability of the model (Fig.~\ref{scale}). It features very good speed-up up to hundreds of processes when deployed in parallel. These results are for so-called ``strong scaling'' where the number of processes is increased while the total problem size, in our case the number of nucleotides, is kept constant. 
The scaling tests were performed on ARCHER, a Cray XC30 supercomputer with 4920 compute nodes, each consisting of two 2.7 GHz 12-core Intel Ivy Bridge processors and Aries Interconnect (Dragonfly topology).
Two different benchmarks were investigated. They consisted both of linear, double-stranded DNA strands of a length of $600$ bp each. The strands were initialised as a regular array of $10\times10 $ or $40\times40$ strands, respectively to form a total system of $60$ kbp and $960$ kbp. The daily simulation times were derived from the loop timings of runs with $30,000$ timesteps ($60$ kbp) and $10,000$ timesteps ($960$ kbp) and were compared with those of a run with 24 processes (MPI-tasks), corresponding to one fully occupied node on ARCHER. We made use of the ``shift'' load-balancing algorithm in LAMMPS, which re-positions the cutting planes between the single processes in order to mitigate a potential load imbalance between the individual processes (further details and full input files are available upon request).

For the smaller problem size of $60$ kbp we observe a parallel efficiency of about 50\% at 512 MPI-tasks, allowing to run for about $2$ ms per day. More processes do not lead to a further speedup and the parallel efficiency decreases rapidly due to the relatively small number of ``atoms'' per process (LAMMPS requires several hundred atoms per process to show good scaling behaviour).
The larger benchmark of $960$ kbp shows a parallel efficiency of about $50\%$ at $2048$ MPI-tasks, which permits simulation times of about $0.4$ ms per day. Compared to the smaller benchmark the performance degrades more slowly in this case, making simulation times of up to $1$ ms per day at $8192$ MPI-tasks feasible. These results strongly encourage its use on a larger scale. 

Other existing models~\cite{Ouldridge2011a,Hinckley2013} might therefore be more suitable for studies of DNA-DNA hybridisation leading to DNA origami and synthetic DNA assemblies. The model we presented here might instead be more apt to study denaturation, supercoiling and DNA-protein interactions as previously discussed.

Finally, exploiting the ability of LAMMPS to function as a library coupled to external programs, we aim to design systems in which ATP-driven proteins interact with the model dsDNA. This paves the way to the attractive avenue of molecular dynamics simulation of large-scale out-of-equilibrium and biologically inspired systems, which are appealing to a broad range of researchers.    

\section{Conclusions}

In summary, we have introduced a coarse-grained single-nucleotide model for dsDNA, which can be readily implemented in computationally efficient and parallelised engines. We tuned the model in order to reproduce the crucial physical features of dsDNA such as bending and torsional rigidities. We then tested our model by simulating single-molecule experiments so as to independently check the parameterisation and the response of our model to external manipulation. Finally, we studied denaturation and the dynamics of supercoiled DNA. We have shown that this implementation can comfortably reach length and time scales that are relevant to both single molecule and biological experiments, therefore making our model interesting for applications. In the future we intend to refine this model and to extend it in order to study biologically-inspired out-of-equilibrium scenarios.      
\subsubsection*{Acknowledgement} 
This work was funded by ERC (Consolidator Grant 648050 THREEDCELLPHYSICS). OH acknowledges support from the EPSRC Reseach Software Engineer Fellowship Scheme (EP/N019180/1). This work used the ARCHER UK National Supercomputing Service. Y.A.G.F. acknowledges support from \-CONACyT\- PhD Grant 384582.

\appendixtitleon
\appendixtitletocon
\begin{appendices}

  \section{Details of the model}
  \label{apendixa}

The dynamics of the system are evolved using the Large-scale Atomic/Molecular Massively Parallel Simulator (LAMMPS). The position of the $i$th atom in the system, $\bm{x}_i$, obeys the Langevin equation
\begin{equation}
	m\dfrac{d^2 {\bm x}_i}{dt^2} = -\gamma  \frac{d\mathbf{x}_i}{dt} - {\bm \nabla} U_i + {\bm \eta}_i 
\end{equation}
where $\gamma$ is the friction coefficient and ${\bm \eta}_i$ is a stochastic noise term which satisfies $\langle \eta_{\alpha}(t)\eta_{\beta}(t') \rangle= 2 \gamma k_BT \delta_{\alpha\beta}\delta(t-t')$. The term ${\bm \nabla} U_i$ is the gradient of the total potential $U_i$ affecting bead $i$, whose contributions are described below. 

\subsection{Bonded interactions}
The interactions between two consecutive beads in the same strand $i$ and $i + 1$ are modelled by the Finite Extensible Non-linear Elastic (FENE) potential:
\begin{equation}
U_{\rm bb}(r) = \begin{cases} 
-\dfrac{K_{1}R^{2}_{0}}{2} \; \text{ln}\left[ 1-\left( \dfrac{r}{R_{0}} \right) ^{2} \right]  & \text{if } r<R_{0} \\
\infty , & \text{if } r\geq R_{0}.
\end{cases}
\end{equation}
where $R_{0}$ is the maximum bond length, $K_{1}$ is the spring constant and $r$ is the Euclidean distance between bead $i$ and bead $i+1$. When summed to the Lennard-Jones potential (acting between any two beads), the minimum of this potential is located at $r_{\rm min}=0.96$ $\sigma_s$. 

The ``hydrogen bond'' is mimicked by a truncated harmonic potential between the patches along the two strands ($i$ and $i^\prime$). This potential reads
\begin{equation}
U_{\rm hb}(r) = \dfrac{K_{2}}{2(r_{0}-r_{c})^{2}}\left[ (r-r_{0})^{2}-(r_{c}-r_{0})^{2}  \right]
\end{equation}
if $r\leq r_{c}$, and $0$ otherwise. Here $r$ represents the distance between patches $i$ and $i^{\prime}$, $r_{0}$ the equilibrium bond distance, $K_{2}$ the spring constant, and $r_{c}$ is the critical distance above which the bond breaks. The minimum of this potential is located at $r=r_{0}$.

\subsection{Non-Bonded interactions}
The excluded volume between beads is modelled via a truncated and shifted Lennard-Jones (LJ) potential. This potential acts between all possible pairs of beads so as to avoiding overlapping, and has the following form: 
\begin{equation}
U_{\rm LJ}(r) = 4\epsilon \left[ 
\left( \frac{\sigma_s}{r} \right)^{12} 
- \left( \frac{\sigma_s}{r} \right)^{6} + \frac{1}{4} \right] ,
\end{equation}
for $r<2^{1/6}\sigma_s$, and $0$ otherwise. Here $\sigma_s$ represents the diameter of a spherical bead, $\epsilon$ parametrises the strength of the repulsion and $r$ is the Euclidean distance between the beads. The minimum of this potential is located at $r=r_{c}=2^{1/6}\sigma_s$. 

The dihedral interaction which regulates the handedness of the chain is given by:
\begin{equation}
U_{\rm dihedral}(\phi) = K_{3} [1+\text{cos}(\phi-d)],
\end{equation}
where $\phi$ is the angle between planes formed by the triplets described in Sec.~\ref{model} and $d$ is a phase angle related to the equilibrium helical pitch.


The stacking of consecutive base-pairs is set by a combination of a Morse potential constraining the distance between consecutive patches
\begin{equation}
U_{\rm morse}(r) = K_{4}[1-e^{-\lambda(r-r_{0})}]^{2}.
\end{equation}
where $r_0$ is the equilibrium distance. A stiff harmonic potential setting the angle $\alpha$ between the tangent along one strand and the vector joining a bead to its patch, imposes the planarity between consecutive patches.
\begin{equation}
U_{\rm harmonic}(\alpha) = \frac{K_{5}}{2}(\alpha - \alpha_{0})^{2}.
\label{harm}
\end{equation}
As described in Sec.~\ref{model} the minimum of this potential is set to $\alpha_{0}=90\degree$.

Finally, the bending rigidity is given by a potential on the angle $\theta$ formed by three consecutive patches that reads
\begin{equation}
U_{\rm bending}(\theta) = K_{6} [1+\text{cos}(\theta)].
\end{equation}

The parameters for each potential are reported in simulation units in Table~\ref{parameters}.

\begin{table}[htb]
	\centering
	\begin{tabular}{|l|l|}
		\hline
		{\bf Interaction}         &   {\bf Parameters}\\
		\hline
		\hline
		
		Backbone: $U_{\rm bb}$        & $K_{1}=30$, $R_{0}=0.6825$,\\
		& $\epsilon=1$ and $\sigma_s=0.4430$\\
		\hline
		Hydrogen bond: $U_{\rm hb}$   & $K_{2}=6$, $r_{0}=0$\\
		& and $r_{c}=0.3$    \\
		\hline
		Steric: $U_{\rm LJ}$          & $\epsilon=1$ and $\sigma_s=1$  \\
		\hline
		Dihedral: $U_{\rm dihedral}$  & $K_{3}=50$, $n=1$,\\
		& and $d=-144\degree$ \\
		\hline
		Morse: $U_{\rm morse}$        & $K_{4}=30$, $\lambda=8$\\
		& and $r_{0}=0.34$\\
		\hline  
		Planarity: $U_{\rm harmonic}$ & $K_{5}=200$ and $\alpha_{0}=90\degree$\\
		\hline
		Bending: $U_{\rm bending}$    & $K_{6}=52$\\
		\hline
	\end{tabular}
	\caption{Parameter values in the model and expressed in simulation units.}
	\label{parameters}
\end{table}
\hspace{1cm}

\subsection{Simulation units}

Mapping the simulation units to physical ones can be done by setting the fundamental units: distance, energy and time. These are shown in Table~\ref{units}. The chosen system of reference is a bath at room temperature $T=300$ K and with the viscosity of water $\eta = 1$ cP.

\begin{table}[h]
\centering
	  	\begin{tabular}{|l|l|l|}
	  	\hline
		{\bf Parameter} & {\bf Experimental units}\\
	  	\hline
	  	\hline
	  	Distance ($\sigma_s$)                        & $1$ nm $\simeq$ $3$ bp \\
	  	\hline
	  	Energy ($\epsilon=k_{B}T$)                 & $4.1419\times 10^{-21}$ J\\
	  	\hline
	  	Force ($F=\epsilon/\sigma_s$)        & $4.1419 \times 10^{-12}$ N \\
	  	\hline
	  	Mobility ($\mu=1/(3\pi\eta \sigma_s$)) & $1.06 \times 10^{11}$ m/Ns  \\
	  	\hline
	  	Diffusion ($D=\mu k_{B}T$)                 &  $4.39 \times 10^{-10}$ m$^{2}$/s\\
	  	\hline
	    Time ($\tau_{Br}=\sigma_s^2/D$)                 &  $2.28 \times 10^{-9}$ s\\
	  	\hline	  	
	  	\end{tabular}
\caption{Mapping between simulation and physical units.}
\label{units}
\end{table}
	  
Finally, the numerical integration is performed in an NVT ensemble by a standard velocity-Verlet algorithm with integration time-step
\begin{equation}
\Delta t = 0.005 \tau_{Br}.
\end{equation}

\section{Computing the torsional persistence length}\label{ap_tors}

To obtain the torsional properties of the DNA molecule described in Sec.~\ref{tors_sec} we consider a discrete elastic rod. As described in Ref.~\cite{Brackley2014a}, Eq.~\ref{tors_eq1} is an integral over the rate of rotation of the Darboux frame (or material frame) of reference with respect to the distance along the rod.  We first find the discrete approximation to this in terms of the Euler angles $\alpha_n,\beta_n,\gamma_n$ which describe the rotation which generates the frame at segment $n+1$ from that at segment $n$. To do this we make the approximation that the step size between segments is constant and denote it $a$; this gives
\[
\frac{E_{\rm tors}}{k_BT}=\frac{C}{a} [ 1 - \cos(\alpha_n+\gamma_n) ],
\]
where twist angle between the frames is given by $\alpha_n+\gamma_n$, so the total angle between $m$ consecutive beads is given by $\Omega(m)=\sum_{n=1}^{m} (\alpha_n+\gamma_n)$. An appropriate measure of the thermal fluctuations about the equilibrium twist is given by the mean of the cosine of this angle; since this quantity will decrease with $m$, and we identify the decay constant as the torsional persistence length $\langle \cos \Omega(m) \rangle =e^{-ma/l_{\tau}}$.
The ensemble average is found in the usual way by taking the integral over the phase space of the system; in the small $a$ limit this gives
\[
l_\tau = 2C.
\]
This is the case for an elastic rod; for the DNA molecule, the non-zero equilibrium twist between each base-pair will appear in the energy functional, so this must be subtracted from $\Omega(m)$ so that the ensemble average is a simple exponential decay. The Darboux frame at each DNA base-pair is given by the tangent vector, and the normal vector defined as the projection of the vector connecting the two beads onto the plane perpendicular to the tangent.

\section{Closure Procedure for Linear DNA}
\label{closure_lk}
\begin{figure}
	\centering
	\includegraphics[width=0.45\textwidth]{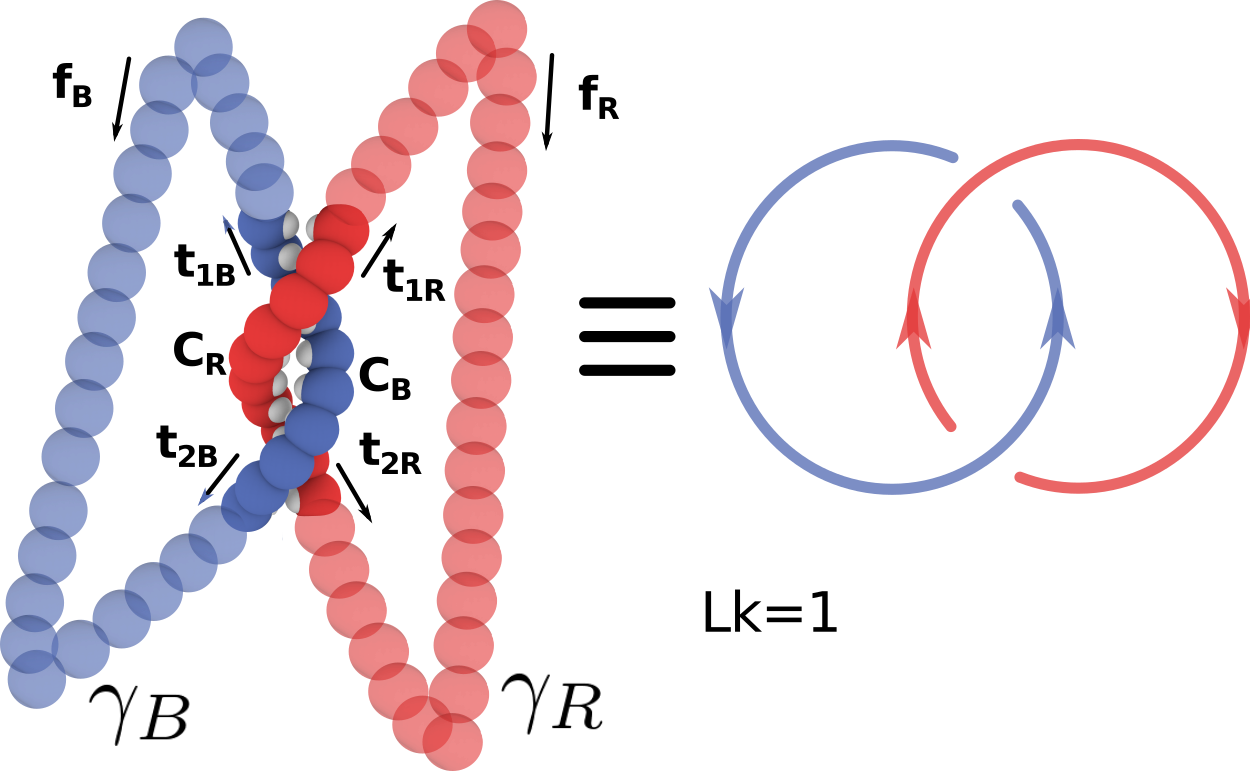}
	\caption{The ``closure'' procedure can be performed on a pair of linear open curves to construct a closed pair whose linking number can be formally defined through the Gauss' integral (see eq.~\eqref{eq:lkint}). In this case the curves are linked once. See text for further details.}
	\label{fig:clos}	
\end{figure}

In this section we review the procedure to compute the linking number of an open segment of dsDNA. For clarity we report a schematic in Fig.~\ref{fig:clos}.
Given two curves $C_R$ and $C_B$ mapping the interval $I = \left[0:1\right] \rightarrow \mathbb{R}^3$, it is possible to formally compute their linking number only if closed, \emph{i.e.} $C_R(0)=C_R(1)$ and $C_B(0)=C_B(1)$. For a linear open segment of dsDNA, a pair of closed strands can be defined by considering the vectors tangent to the terminal pair of beads of the two single strands forming the dsDNA segment and extending the curves away from the pair of strands. Reached a certain distance by following, for instance, $\bm{t}_{1R}$ and $\bm{t}_{2R}$, one can close the contour by defining a vector $\bm{f}_R$ that joins the two new terminal beads (see Fig.~\ref{fig:clos}). By following this procedure one can finally construct a pair of closed oriented curves $\gamma_R$ and $\gamma_B$, for instance ``stitching'' $C_R$, $\bm{t}_{1R}$,$\bm{f}_R$,$-\bm{t}_{2R}$, and similarly for the blue curve.
Their linking number can be computed through the numerical evaluation of the double integral
\begin{equation}
	Lk(C_R,C_B) = \dfrac{1}{4\pi} \int_{\gamma_{R}} \int_{\gamma_B} \dfrac{|\bm{r}_R - \bm{r}_B|}{|\bm{r}_{R} - \bm{r}_B|^3} \cdot \left(d\bm{r}_R \times d\bm{r}_B\right), \label{eq:lkint}
\end{equation}
 where $\bm{r}_{R}$ and $\bm{r}_{B}$ are the vectors defining the position of the segments along the curves $\gamma_R$ and $\gamma_B$, respectively.
 If the centreline running through the pair of curves has no self-intersections (null writhe) then the linking number is equal to the twist. 
It is also worth mentioning that tightly wound curves, such as those obtained from dsDNA configurations, can lead to imprecise numerical evaluation of the integrals in eq.~\eqref{eq:lkint}. In fact, the computation of $Lk$ can become unreliable when $|\bm{r}_R - \bm{r}_B| \simeq d\bm{r}_R \simeq d\bm{r}_B$. The numerical evaluation can be arbitrarily improved by replacing the DNA backbones by contours more finely interspersed with points, \emph{i.e.} enhancing the resolution of the integral by decreasing the infinitesimal element $d\bm{r}$. Clearly, this can slow down the computation of $Lk$. We found a good compromise between precision and speed by adding three intermediate points every pair of beads for which we consistently measured the correct linking number during topology-preserving simulations (for instance by considering circular dsDNA).  

\end{appendices}

\footnotesize{
\bibliography{dsDNA,lammps,supercoil_temp,bibliography1} 
\bibliographystyle{rsc} 
}

\end{document}